\documentclass[aps,prx,twocolumn,superscriptaddress,notitlepage,nofootinbib]{revtex4-1}
\usepackage{graphicx}
\usepackage{upgreek}
\usepackage{breakcites}
\usepackage[linktoc=page,colorlinks,urlcolor=blue,citecolor=blue,linkcolor=blue]{hyperref}
\usepackage{amssymb,amsmath}
\usepackage{graphicx}
\usepackage{natbib}
\usepackage{mathrsfs}
\usepackage{overpic}
\usepackage{wrapfig}
\usepackage{xcolor}
\usepackage{textcomp}
\usepackage{comment}
\usepackage{multirow}

\usepackage{gensymb}
\usepackage{ulem}
\usepackage{color}
\usepackage[letterpaper,textwidth=7in,top=.75in,bottom=.75in]{geometry}
\newcommand{\huPhys}{Department of Physics, Harvard University, Cambridge, MA 02138, USA}
\newcommand{\CfA}{Harvard-Smithsonian Center for Astrophysics, Cambridge, MA 02138, USA}
\newcommand{\cbs}{Center for Brain Science, Harvard University, Cambridge, MA 02138, USA}
\newcommand{\MITRE} {The MITRE Corporation, Bedford, MA 01730, USA}
\newcommand{\Sandia}{Sandia National Laboratories, Albuquerque, NM 87123, USA}
\newcommand{\SEAS} {John A. Paulson School of Engineering and Applied Sciences, Harvard University, Cambridge, Massachusetts 02138, United States}
\newcommand{\MaryPhys} {Department of Physics, University of Maryland, College Park, MD 20740, USA}
\newcommand{\MaryECE} {Department of Computer and Electrical Engineering, University of Maryland, College Park, MD 20740, USA}
\newcommand{\MaryQTC} {Quantum Technology Center, University of Maryland, College Park, MD 20740, USA}

\begin{document}

\title{Magnetic Field Fingerprinting of Integrated Circuit Activity with a Quantum Diamond Microscope}
\date{\today}

\author{Matthew J. Turner} 
\affiliation{\huPhys} 
\affiliation{\cbs}

\author{Nicholas Langellier} 
\affiliation{\huPhys} 
\affiliation{\CfA}

\author{Rachel Bainbridge} 
\affiliation{\MITRE}

\author{Dan Walters} 
\affiliation{\MITRE}

\author{Srujan Meesala} 
\altaffiliation[Current Affiliation: ]{Thomas J. Watson, Sr. Laboratory of Applied Physics, California Institute of Technology, Pasadena, California 91125, USA}
\affiliation{\huPhys} 

\author{Thomas M. Babinec}  
\affiliation{\SEAS}

\author{Pauli Kehayias} 
\affiliation{\Sandia}

\author{Amir Yacoby} 
\affiliation{\huPhys} 
\affiliation{\SEAS}

\author{Evelyn Hu} 
\affiliation{\SEAS}

\author{Marko Lon\v{c}ar} 
\affiliation{\SEAS}

\author{Ronald L. Walsworth}  
\affiliation{\huPhys} 
\affiliation{\cbs} 
\affiliation{\CfA} 
\affiliation{\MaryPhys} 
\affiliation{\MaryECE} 
\affiliation{\MaryQTC}

\author{Edlyn V. Levine} 
\email[Corresponding Author: ] {edlynlevine@fas.harvard.edu}
\affiliation{\huPhys} 
\affiliation{\MITRE}

\begin{abstract}
Current density distributions in active integrated circuits (ICs) result in patterns of magnetic fields that contain structural and functional information about the IC. Magnetic fields pass through standard materials used by the semiconductor industry and provide a powerful means to fingerprint IC activity for security and failure analysis applications. Here, we demonstrate high spatial resolution, wide field-of-view, vector magnetic field imaging of static (DC) magnetic field emanations from an IC in different active states using a Quantum Diamond Microscope (QDM). The QDM employs a dense layer of fluorescent nitrogen-vacancy (NV) quantum defects near the surface of a transparent diamond substrate placed on the IC to image magnetic fields.  We show that QDM imaging achieves simultaneous $\sim10$ $\mu$m resolution of all three vector magnetic field components over the 3.7 mm $\times$ 3.7 mm field-of-view of the diamond. We study activity arising from spatially-dependent current flow in both intact and decapsulated field-programmable gate arrays (FPGAs); and find that QDM images can determine pre-programmed IC active states with high fidelity using machine-learning classification methods.
\end{abstract}

\maketitle
\section{Introduction}

Securing integrated circuits (ICs) against manufacturing flaws, hardware attacks, and software attacks is of vital importance to the semiconductor industry \cite{rostami2014primer}.  Hardware attacks often modify the physical layout of an integrated circuit, thereby changing its function. This type of attack can occur at any stage of the globalized semiconductor supply chain, and can range from insertion of malicious Trojan circuitry during the design and fabrication stages \cite{tehranipoor2010trustworthy}, to modification or counterfeiting during packaging and distribution stages \cite{semiconductor2013winning}. Horizontal integration of the industry has led to contracting of IC fabrication, packaging, and testing to offshore facilities, resulting in a reduction of secure oversight and quality control \cite{hoeper2017dsb}. Additional growth of the secondhand electronics market has led to a drastic increase in counterfeit ICs \cite{guin2014counterfeit}. Detection of IC tampering or counterfeiting has consequently become essential to ensure hardware can be trusted. Similar issues affect quality control of unintended manufacturing flaws.

\begin{figure*}[htbp]
\begin{center}
\begin{overpic}[width=0.95\textwidth]{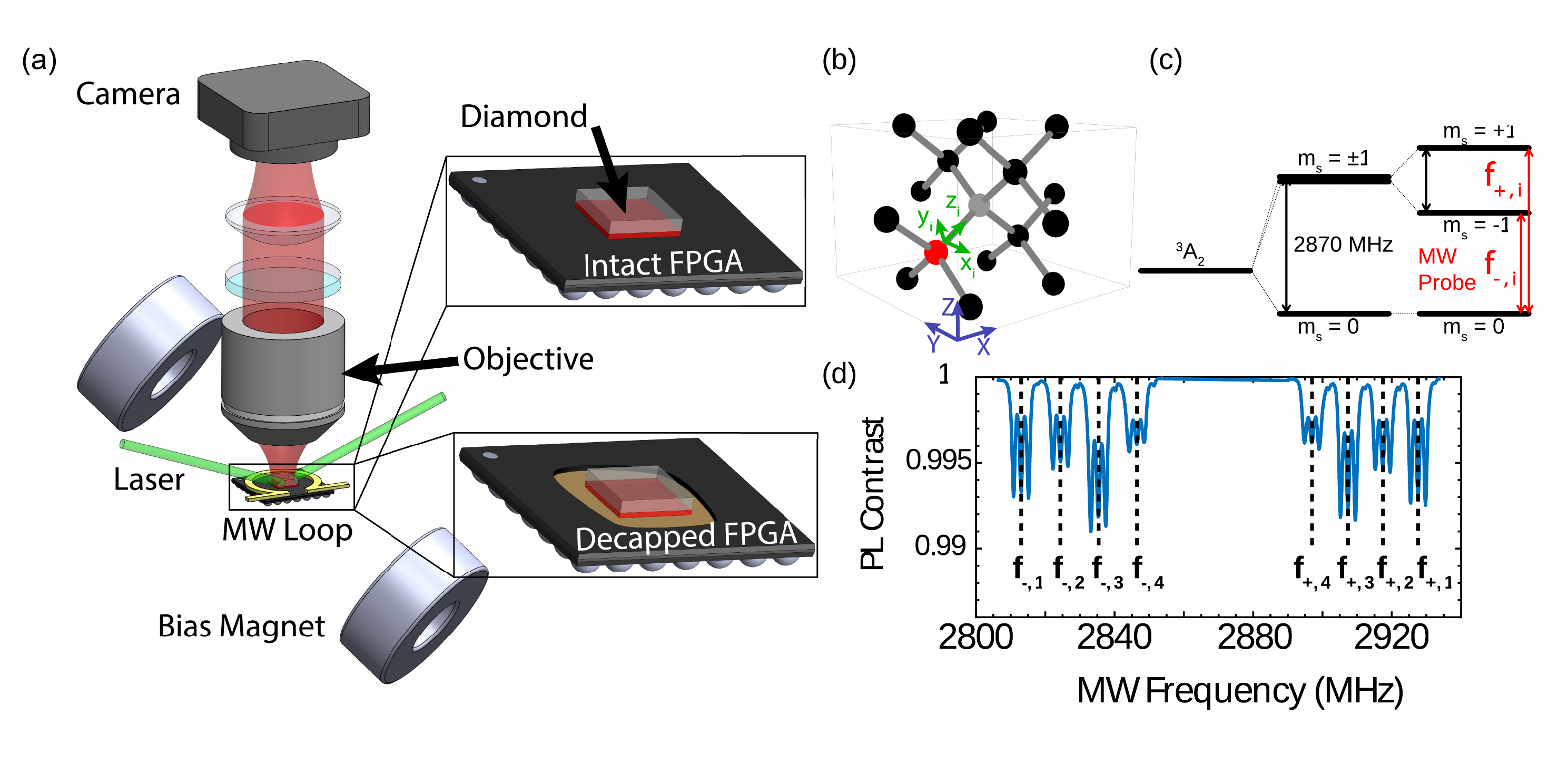}
\end{overpic}
\end{center}
\caption{\label{Experiment}
(a) Schematic of the Quantum Diamond Microscope (QDM) experimental setup with insets showing the diamond in contact with intact and decapsulated FPGAs. The diamond is positioned such that the NV layer is in direct contact with the FPGA, as indicated by the red layer in the insets.  (b) Diamond crystal lattice with the nitrogen (red) vacancy (grey) defect. Lab frame coordinates ($X$,$Y$,$Z$) and NV frame coordinates for a single defect ($x$,$y$,$z$) are shown. (c) The ground state energy level diagram for an NV with fine structure and Zeeman splitting. (d) Example ODMR spectral data for an applied bias field of ($B_X$,$B_Y$,$B_Z$) = (2.04 1.57 0.65) mT, showing resonance frequencies of $f_{\pm,i}$ with $i=1,2,3,4$ indicating each of the four NV axes. Hyperfine interactions between the NV$^-$ electrons and the spin-1 $^{14}$N nucleus result in splitting of each NV resonance into three lines.}
\end{figure*}

Magnetic field emanations from ICs afford a powerful means for non-destructive physical testing. Magnetic fields are generated by current densities in ICs resulting from power and clock distribution networks, input/output lines, word and bit lines, and switching transistors. These currents are present in all operating logic and memory chips and can be leveraged for studying the operational behavior of an IC during task execution. In general, the resulting IC magnetic fields pass through many standard IC materials, and will vary spatially and temporally in ways that correlate with both IC architecture and operational state.  Thus, combined high-resolution and wide-field-of-view mapping of magnetic fields may yield simultaneous structural and functional information, and may be suitable for identification of malicious circuitry or Trojans \cite{balasch2015electromagnetic, gaudestad2014magnetic}, counterfeit detection \cite{tagro2017innovative}, fault detection \cite{orozco2019magnetic,campbell1993internal}, and manufacturing flaws \cite{huston1992reliability}. However, leveraging magnetic field emanations is challenging due to the tremendous complexity of circuits integrating billions of transistors of minimum feature sizes down to tens of nanometers, with interconnects distributed across multiple levels of metallization  \cite{IndustryVision2017}. Multi-layered metal interconnects and three-dimensional stacking give rise to complex magnetic field patterns that are difficult to invert; and large stand-off distances of magnetometers reduce amplitudes of magnetic fields and spatial resolution \cite{roth1989using}. 

In this paper, we demonstrate how these challenges can be approached using a Quantum Diamond Microscope (QDM) \cite{qdmEdlyn, QDM1ggg,taylor2008high} augmented with machine-learning classification techniques. With the QDM, we perform simultaneous wide field-of-view, high spatial resolution, vector magnetic field imaging of an operational field-programmable gate array (FPGA). FPGAs are configurable ICs that are commonly used for diverse electronics applications. Systematic and controlled variation of the circuit activity in the FPGA generates complex magnetic field patterns, which we image with the QDM. The QDM employs a dense surface layer of fluorescent nitrogen-vacancy (NV) quantum defects in a macroscopic diamond substrate placed on the IC under ambient conditions.

We use the QDM to image magnetic fields from both decapsulated (decapped) and through-package (intact) FPGAs under operational conditions using continuous wave (CW) optically detected magnetic resonance (ODMR) NV spectroscopy. For the decapped FPGA, our measurements yield magnetic field maps that are distinguishable between operational states over approximately a 4~mm $\times$ 4~mm field-of-view with a 20~nT noise floor, and $\sim$ 10~$\mu m$ magnetic field spatial resolution, limited by the thickness of the NV surface layer in the diamond and the distance to the nearest metal layer. For the intact FPGA, the QDM measurements provide magnetic field maps with a similar field-of-view, 2~nT noise floor, and $\sim$ 500~$\mu m$ magnetic field spatial resolution, limited by the stand-off distance between the NV layer and the FPGA current sources. In particular, we find that operational states of the intact FPGA are distinguishable in the QDM images, even with the diminished magnetic field amplitude and spatial resolution that arise from the large stand-off between the diamond and the IC die. We use machine learning methods to demonstrate FPGA operational state classification via magnetic field pattern correlation for both decapped and intact FPGA QDM images. This result provides an initial demonstration of functional IC characterization that can be leveraged to fingerprint IC activity.

To date, QDM magnetic field imaging has been used to measure microscopic current and magnetization distributions from a wide variety of sources in both the physical and life sciences \cite{pham2011magnetic,ku2019imaging,barryNeurons,jfRochICs, hollenbergFilms,nv_bacteria,Fu1089}. Complementary to existing scanning techniques for characterizing IC magnetic field emanations, which include wire loops \cite{de2010electromagnetic}, probe antennas \cite{sauvage2009electromagnetic}, magnetic force microscopy \cite{campbell1993internal}, SQUID magnetometers \cite{gaudestad2014magnetic}, and vapor cell magnetometers \cite{horsley2015widefield}, the QDM provides simultaneous high-resolution (micron-scale) and wide-field (millimeter-scale) vector magnetic imaging. This capability allows for monitoring of transient behavior over sequential measurements of a magnetic field, providing a means to study correlations in signal patterns that can evolve more quickly than a single-sensor scan time. With these distinctive advantages, the QDM-based technique affords a promising, novel means for non-destructive physical testing of ICs.

\section{Experimental Design}

\subsection{QDM Experimental Setup}
A schematic of the QDM is shown in Fig.~\ref{Experiment}(a). The magnetic field sensor consists of a 4 mm $\times$ 4 mm $\times$ 0.5 mm diamond substrate with a 13 $\mu$m surface layer of NV centers. The diamond is placed directly on the IC with the NV layer in contact with the IC surface. The diamond is grown by Element Six Ltd. to have an isotopically pure NV layer consisting of [$^{12}$C] $\sim 99.999 \%$, [$^{14}$N] $\sim$ 27 ppm, and [NV$^\text{-}$] $\sim$ 2 ppm.  Light from a 532 nm, CW laser (Lighthouse Photonics Sprout-H-10W) optically addresses the NV layer with beam power of about 500 mW uniformly distributed over the 4 mm $\times$ 4 mm NV layer. A flat-top beam shaping element (Eksma Optics GTH-5-250-4-VIS) and a cylindrical lens (Thorlabs LJ1558RM-A) create a rectangular beam profile (6 mm $\times$ 6 mm)  incident on the top face of the diamond at a sufficiently shallow angle of incidence (4$^\circ$) relative to the top diamond surface to illuminate the entire NV layer. NV fluorescence is collected with a low magnification objective (Olympus UPlanFL N 4x 0.13 NA) to interrogate a large field-of-view of about 3.7 mm $\times$ 3.7 mm. The fluorescence is filtered with a 633 nm longpass filter (Semrock LP02-633RU-25) and imaged onto a CMOS camera (Basler acA1920-155um).  Resulting CW ODMR data is transferred to a computer where it is processed and analyzed with custom software utilizing LabVIEW and MATLAB.

A pair of  5~cm diameter SmCo permanent magnets (Super Magnet Man) is placed on opposing sides of the diamond to apply a uniform bias magnetic field (bias field) of $\mathbf{B}_0 = $ ($B_X$, $B_Y$, $B_Z$) = (2.04, 1.57, 0.65)~mT. $X$, $Y$, $Z$ are the laboratory frame Cartesian coordinates with the X-Y plane defined as the surface of the diamond in Fig.~\ref{Experiment}(b). $\mathbf{B}_0$ induces a $\pm g_e\mu_B\mathbf{B}_0\cdot\mathbf{n}$ Zeeman splitting of the spin triplet NV $m= 1$ and $m=-1$ ground states along each of four tetrahedrally defined NV symmetry axes, $\mathbf{n}$, with Land\'{e} g-factor $g_e$, and Bohr magneton $\mu_B$. The hyperfine interaction between the NV and the $^{14}$N isotope nuclear spin (I=1) results in an additional triplet level splitting. The four symmetry axes of the NV, shown in Fig.~\ref{Experiment}(b), are leveraged for vector magnetic field imaging using $\mathbf{B}_0$ projection onto all four NV axes \cite{schloss2018simultaneous}. The ground state energy level diagram of a single NV axis is depicted in Fig.~\ref{Experiment}(c), neglecting hyperfine structure. 

A 6 mm diameter copper wire loop made from 320 $\mu$m diameter magnet wire delivers 1 W of GHz-frequency microwave (MW) fields (TPI-1001-B and amplified with a Mini-Circuits ZHL-16W-43S+ amplifier) to drive the NV electronic spin transitions, $m_s=0 \leftrightarrow -1$ or $m_s=0 \leftrightarrow +1$, denoted by $f_{-,i}$ and $f_{+,i}$, respectively, see Fig.~\ref{Experiment}(c). The MW field is modulated on and off through the use of a solid-state switch (ZASWA-2-50DRA+) controlled by a DAQ (NI-USB 6259) and synchronized with the frame acquisition of the camera to correct for laser intensity fluctuations and drift. 

The intensity of optically induced NV fluorescence decreases for MW fields on resonance with one of the spin transition energies. This decrease results from the $m=\pm 1$ spin-selectivity of the non-optical, intersystem crossing (ISC) mediated decay pathway for optically excited NVs \cite{marcus_review}. The resonance frequencies between NV ground-state sublevels are determined from the ground-state Hamiltonian
\begin{align}
\begin{split}
H/h = & \left( D(T) + M_{z} \right) S_{z}^2 + \gamma \left(B_x S_x + B_y S_y + B_z S_z \right) \\
& + M_{x} \left( S_{y}^2 - S_{x}^2 \right) + M_{y} \left( S_{x} S_{y} + S_{y} S_{x} \right)
\label{nvHamiltonian}
\end{split} 
\end{align}  
for the projection of $\mathbf{B}_0$ along a single NV axis, where $h$ is Planck's constant, $D(T) \approx 2870$~MHz is the temperature dependent zero field splitting, $T$ is the temperature, $S_k$ are the dimensionless spin-1 Pauli operators, $\gamma=2.803\times10^4$~MHz/T is the NV gyromagnetic ratio, $B_k$ are the components of $\mathbf{B}_0$ in the NV frame, and $M_k$ are crystal stress terms \cite{StrainPaper}. Cartesian coordinates $k = x,y,z$ are defined in the NV frame with $z$ along the selected NV axis, see Fig.~\ref{Experiment}(b). The contribution of the hyperfine interaction between the NV and $^{14}N$ nuclear spin is treated as a constant, 2.158~MHz energy level splitting and is not shown explicitly in Eqn.~\ref{nvHamiltonian}. Sweeping the frequency of the applied MW fields across the range of resonant frequencies and collecting the NV fluorescence results in an optically detected magnetic resonance (ODMR) spectrum. Fig.~\ref{Experiment}(d) depicts the resulting ODMR measurements for a bias field alignment where each NV axis experiences a different projection of the bias magnetic field. 

Continuous-wave (CW) ODMR is used to image DC FPGA magnetic fields. CW ODMR leverages continuous application of the laser and MW field. This approach yields wide field-of-view images with high spatial resolution and good magnetic field sensitivity, while minimally perturbing the sample under study \cite{qdmEdlyn,QDM1ggg}. A diamond with sufficiently low $M_z$ inhomogeneity across the field-of-view is used to minimize degradation of performance \cite{StrainPaper}. Further suppression of strain contributions is achieved with application of the static bias field, $\mathbf{B}_0$. Thus, the $M_x$ and $M_y$ terms in Eqn.~\ref{nvHamiltonian} are negligible \cite{P1DQ, StrainPaper}. The ground state Hamiltonian along a single NV axis reduces to 

\begin{equation}
H/h \approx (D(T)+M_{z}) S_{z}^2 + \gamma B_{z} S_{z} + \gamma B_{x} S_{x} + \gamma B_{y} S_{y} ,
\end{equation}
and is used to determine the CW ODMR resonance frequencies for each pixel in a QDM image, and thereby to determine the magnetic field image from the sample FPGA. 

\subsection{IC Preparation, Control, and Layout}
\begin{figure*}[htbp]
\begin{center}
\begin{overpic}[width=0.95\textwidth]{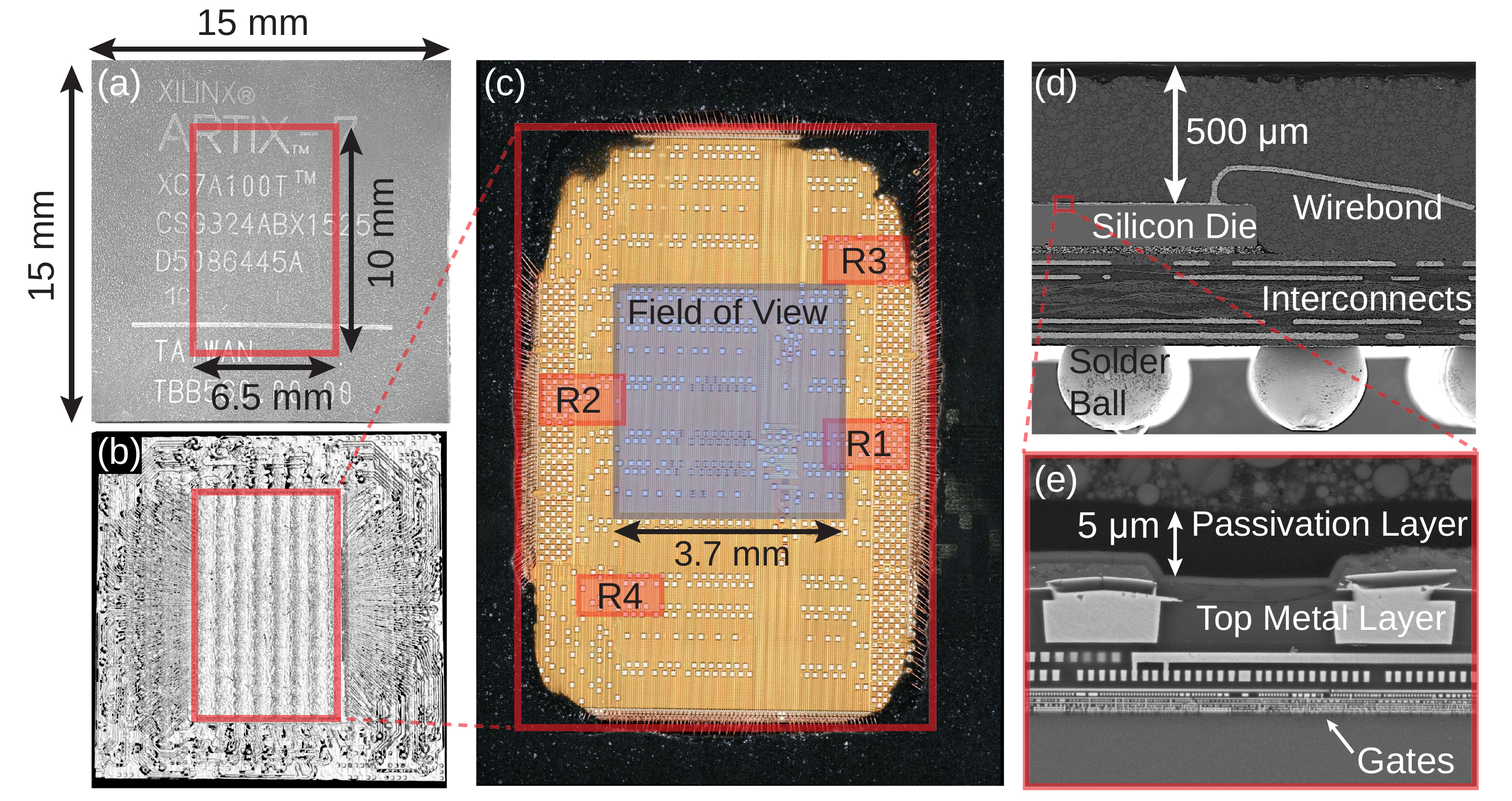}
\end{overpic}
\end{center}
\caption{\label{Artix}
(a) Intact Xilinx 7-series Artix FPGA with die location and dimensions indicated in red (b) X-ray image of the FPGA package determining the position and size of die outlined in red (c) A high resolution image of the decapsulated FPGA with the fixed diamond measurement field-of-view indicated with a blue box, and the location of ring oscillator (RO) clusters indicated by red boxes labeled R1 - R4. (d) Scanning electron microscope (SEM) image of the FPGA cross-section showing the 500~$\mu m$ stand-off distance between the chip die and the top layer. (e) Close-up of the SEM focusing on the metal layers of die. The thickness of the passivation layer is 5~$\mu m$ and sets the minimum stand-off distance for the decapsulated measurements.}
\end{figure*}

The Xilinx 7-series Artix FPGA (XC7A100T-1CSG324C) shown in Fig.~\ref{Artix}(a) and \ref{Artix}(b) is selected for this study. This FPGA is a $15$~mm $\times$ $15$~mm  wirebonded chip, fabricated in the TSMC 28~nm technology node, that has a $\sim 6.5$~mm $\times$ $10$~mm silicon die with eight clock regions.  Digilent Nexys A7 development boards are used to operate and configure the Artix-7 FPGA.

The large current draw and controllable location and size of ring oscillators (ROs) make them ideal functional units for this study \cite{zhang2011ron}. Patterns of ROs are implemented using the Xilinx Vivado Design Suite$^\text{\textregistered}$ to create distinguishable current distributions on the FPGAs for measurement by the QDM. Clusters of three-inverter ROs are synthesized, placed, and routed to four different predefined clock regions on the FPGA, with clear spatial separation and spanning a majority of the die surface as shown in Fig.~\ref{Artix}(c). The clusters consist of variable numbers of ROs allowing for incremental increase or decrease of the current draw at the different locations on the FPGA. The active states of the FPGA are defined by RO clusters implemented in one of the predefined regions, and the idle state is defined as the FPGA powered on with no implemented ROs. The ordering of states during a series of measurements are randomized to reduce susceptibility to systematic noise sources. 

The die of the Artix-7 FPGA is covered by roughly 500~$\mu$m of epoxy resin packaging material, separating the diamond from the die for the intact configuration. This stand-off distance leads to smaller field amplitude at the NV sensor layer and acts as a low-pass filter decreasing the effective QDM spatial resolution of FPGA current sources \cite{roth1989using}. To bring the diamond closer to the die, one of the Artix-7 FPGAs is decapsulated (decapped) using a Nisene JetEtch Pro CuProtect decapsulator (Fig.~\ref{Artix}(c)). This process uses fuming sulfuric and nitric acid to remove the packaging material, exposing the die while leaving the FPGA electrically functional, including preservation of the copper wirebonds.

The structure of the wire-bonded Artix 7 die, shown in Fig.~\ref{Artix}(d), is optimal for studies of patterns of power delivery within the top metal layers of the FPGA. The thickest layers of the metal stack are usually closest to the top side of the package in wirebonded chips. These thick layers are used for power distribution due to their relatively low resistance characteristics compared to the other layers of the IC. Clock distribution networks and inputs/outputs (I/Os) occupy the next thickest layers, and data signals are in the lowest and thinnest metal layers. Prominent magnetic fields from the current densities in the power distribution network are therefore most easily detected with topside access of a wirebonded device. Magnetic field patterns from the lower-level data signals are likely not distinguishable with the present measurements. Note that Fig.~\ref{Artix}(d) reveals large wire interconnects in the package substrate connecting the wirebonds and solder balls. These wires are deeper in the chip and are likely observable as low spatial frequency components in the magnetic field. As will be seen below, static fields from solder balls and other magnetic materials are also observable with the QDM; but can be distinguished from functional current flow by differential RO measurements.

\subsection{Experimental Protocol and Data Analysis}
CW ODMR measurements are taken with the FPGA in both active and idle states. The duration of each measurement is $\sim$~5~min per NV axis for each state (see Supplemental Material \cite{suppl}). Magnetic field contributions from the ROs are determined by subtracting the measured idle-state ODMR frequencies from the measured active-state ODMR frequencies, yielding the overall magnetic field due to the ROs alone. For such measurements, the NV ground-state Hamiltonian is given by:

\begin{align}
\begin{split}
H/h \approx \left(D + \frac{\partial{D}} {\partial{T}} \Delta T + M_{z} \right) S_{z}^2 + \gamma (B_{z}+\Delta B_z) S_{z} \\
+ \gamma (B_{x}+\Delta B_x) S_{x} + \gamma (B_{y}+\Delta B_y) S_{y}  
\end{split}
\end{align}
Here, terms with $\Delta$ originate from the FPGA active states. Following these definitions and treating the off-axis magnetic fields as perturbative, the idle and active-state resonant frequencies for the upper ($f_+$) and lower ($f_-$) transitions of a single NV axis (i) are given by \cite{P1DQ}

\begin{align}
f_{\pm, i, Idle} & \approx (D + M_{z}) +\frac{3 \gamma^2}{2D} \left(B_{x}^2+ B_{y}^2\right) \pm \gamma B_{z}
 \label{nvOff}
\end{align}
and

\begin{align}
\begin{split}
f_{\pm, i, Active}  \approx  \left(D + \frac{\partial{D}} {\partial{T}} \Delta T + M_{z}\right) \\
  +\frac{3 \gamma^2 \left[(B_{x}+\Delta B_x)^2+ (B_{y}+\Delta B_y)^2\right]}{2\left(D+ \frac{\partial{D}} {\partial{T}}  \Delta T + M_z\right)}  \\
  \pm \gamma (B_{z}+\Delta B_z) .
 \label{nvOn}
\end{split}
\end{align}
The desired FPGA state-dependent magnetic field projection on each NV axis, $\Delta B_{z,i}$, and the change in local temperature, $\Delta T$, are given by 
\begin{equation}
\begin{split}
\Delta B_{z,i} & = \frac{1}{{2\gamma}}\left(\Delta f_{+,i}-\Delta f_{-,i}\right),\\
\Delta T & = \frac{1}{2\frac{\partial{D}} {\partial{T}} }\left(\Delta f_{+,i}+\Delta f_{-,i} \right) ,
\end{split}
\label{SolvedFields}
\end{equation}
where $\Delta f_{\pm,i} = f_{\pm,i,Active}-f_{\pm,i,Idle}$. The off-axis magnetic fields of the sample are suppressed by the zero-field splitting; thus terms dependent on $\Delta B_x$ and $\Delta B_y$ are sufficiently small to be neglected in Eqn.~\ref{SolvedFields} (see Supplemental Material \cite{suppl}). Terms dependent on $B_x$, $B_y$, $B_z$, $D$, and $M_z$ are canceled by subtracting the idle resonance frequencies from the active state resonance frequencies. 
Determining the resonance frequencies from all four NV orientations for vector measurements, labeled by $i=1,2,3,4$ in Fig.~\ref{Experiment}(d), enables solving for the vector magnetic field in the lab frame
\begin{equation}
\begin{split}
\Delta B_X & = \frac{\sqrt{3}}{2\sqrt{2}} \left(\Delta B_{z,2} + \Delta B_{z,4}\right),\\
\Delta B_Y & = \frac{\sqrt{3}}{2\sqrt{2}} \left(\Delta B_{z,1}+ \Delta B_{z,3}\right),\\
\Delta B_Z & = \frac{\sqrt{3}}{4} \left[\left( \Delta B_{z,1}- \Delta B_{z,3}\right)-\left( \Delta B_{z,4}- \Delta B_{z,2}\right) \right],
\end{split} 
\label{VectorFields}
\end{equation} 
where $X$, $Y$, $Z$ are recalled to be the laboratory frame shown in Fig.~\ref{Experiment}(b). 

 The ODMR lineshape for NV ensembles is well approximated by a Lorentzian lineshape \cite{P1DQ}. ODMR spectra for vector measurements of a $^{14}$N diamond sample contain 24 resonance features, Fig.~\ref{Experiment}(d) (3 hyperfine features $\times$ 2 electronic spin transitions $\times$ 4 NV axes). The resonance frequencies of Eqn.~\ref{SolvedFields} are extracted from the data by fitting all the Lorentzian parameters for every pixel in the field-of-view \cite{QDM1ggg}. Furthermore, the contrast and linewidth of the resonances are determined, giving additional state dependent information \cite{suppl} which can additionally be used for probing high frequency magnetic fields \cite{MWImaging}. GPU-based fitting algorithms \cite{GPUfit} speed up this computationally intensive fitting and enable rapid analysis of a large number of measurements.

\section{Results}
\subsection{Vector Magnetic Imaging}
\begin{figure*}[htbp]
\begin{center}
\begin{overpic}[width=0.95\textwidth]{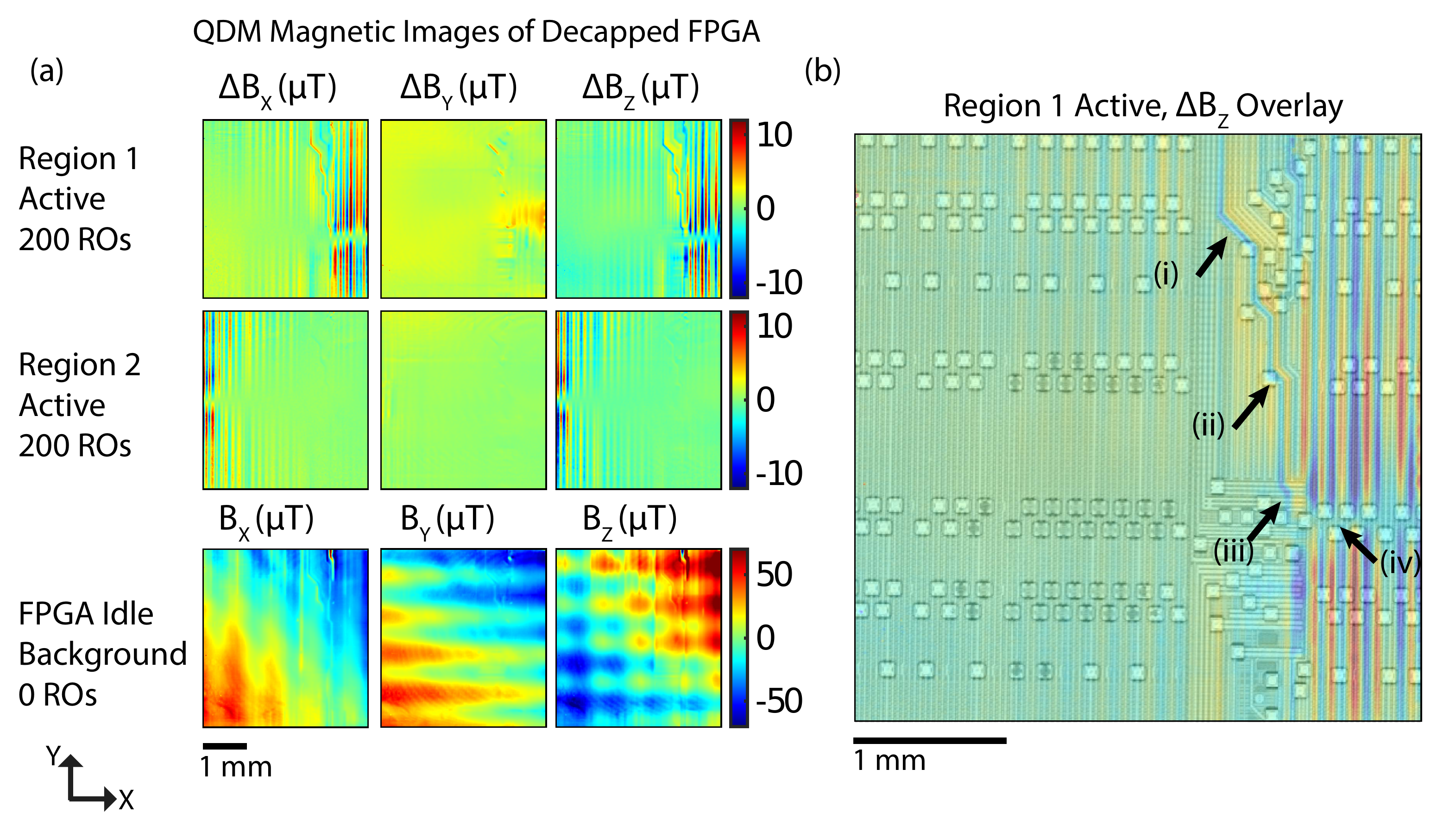}
\end{overpic}
\end{center}
\caption{\label{VectorField}
 (a) QDM vector magnetic field maps of the decapped FPGA for different RO clusters activated in regions, R1 and R2. The location of the 3.7 mm $\times$ 3.7 mm diamond field-of-view is fixed on the FPGA for all magnetic field images (see Fig.~\ref{Artix}(c)). State dependent magnetic field changes ($\Delta B_X, \Delta B_Y, \Delta B_Z$) are calculated by subtracting background idle magnetic field images from active magnetic field images. Wires on the top metal layer are generally oriented in the $Y$ direction yielding prominent $\Delta B_X$ and $\Delta B_Z$ fields. $\Delta B_Y$ magnetic field maps show contributions from deeper sources. Background magnetic field maps of the idle FPGA with 0 ROs show variations of the field from the mean. Several different background fields are evident: a gradient from the bias magnet, distortion of the bias field from the BGA, and background current delivery. (b) $\Delta B_Z$ for 200 ROs in R1 plotted in transparency over a high resolution optical image of circuit die. Regions of interest discussed in the text are indicated by (i),(ii),(iii), and (iv).}
\end{figure*}

Fig.~\ref{VectorField}(a) shows QDM vector magnetic field images measured on the decapsulated FPGA for clusters of $N=200$ ROs in two of the predefined regions indicated in the Vivado floor planner, labeled R1 and R2 in Fig.~\ref{Artix}(c). The vector magnetic field images are derived from CW ODMR measurements using Eqns.~\ref{SolvedFields} and \ref{VectorFields}. Observed maximum magnetic fields are on the order of $\sim$ 15 $\mu$T with a noise floor of $\sim$ 20 nT (see Supplemental Material \cite{suppl}). Spatial variation of the magnetic field is located on the right of the field-of-view for R1 and on the left for R2. This localization corresponds to the positions of R1 and R2 on the Vivado floor planner, indicating that high current densities for power distribution are concentrated to the region of activity on the die. The vector magnetic field measured in the idle state with 0 ROs, shown in the bottom row of Fig.~\ref{VectorField}(a), reveals the structure of the ball grid array (BGA) that connects the FPGA to the Digilent board. The state dependent magnetic fields due to the RO current densities (see Eqn.~\ref{VectorFields}) are thus measured in superposition with the spatially inhomogeneous field resulting from the BGA. 

The presence of a non-zero $B_Y$ component in R1 and R2, as seen in Fig.~\ref{VectorField}(a) indicates contributions to the magnetic field from current density sources that run underneath and perpendicular to the visible traces of the top metal layer. These sources are likely a combination of currents in the lower layers of the metal stack and in the interconnects between the wirebonds and BGA seen in the SEM image in Fig.~\ref{Artix}(d). Discontinuities present in the $B_X$ and $B_Z$ fields indicate a change of the current direction guided by through-silicon vias in the $Z$ direction that connect the different, stacked metal layers. R3 and R4, seen in Fig.~\ref{Artix}(c), are both outside the measurement field-of-view. However, in both cases, state-dependent current is measured in locations corresponding to the direction of current flow in the appropriate location on the die (see Supplemental Information \cite{suppl}). This demonstrates the possibility to determine circuit activity outside of the diamond periphery by observing correlated magnetic fields within the nominal field-of-view.

An optical image of the die through the diamond is used to spatially align the magnetic field measurement with the high resolution optical images taken of the decapsulated chip. Spatial variation of the $B_X$ and $B_Z$ magnetic field components corresponds well with the physical features of the top metal layer. Fig.~\ref{VectorField}(b) shows a zoomed-in overlay of the $B_Z$ field for 200 ROs in R1 with the optical image of the die, demonstrating feature alignment. Distinct features are visible in the fields that correspond to physical structures, including bends in the wires labeled (i) and (ii) in the figure. Some features in the magnetic field map do not correspond to any visible features on the top metal layer, such as the magnetic trace indicated by (iii) or the discontinuity in field direction indicated by (iv). These fields suggest the presence of additional current routing by vias and other structures below the plane of the top metal layer.

\subsection{Single NV-Axis Magnetic Imaging}

Single NV-Axis QDM measurements \cite{QDM1ggg} are used to collect a large data set of magnetic field images from RO clusters for classification. These data are taken by monitoring the outermost ODMR spectral features ($f_{-,1}$ and $f_{+,1}$) with the same bias field in the lab frame, and the laser polarization and MWs optimized for the NV axis being monitored to improve measurement contrast \cite{QDM1ggg}. Projection imaging is useful for large data acquisition due to the speedup in measurement time; however, the vector nature of the field is not captured. The laser polarization and MWs are optimized for the single NV axis being monitored. Measuring only a single pair of resonance features results in a $\sim$ 4$\times$ speed up by reducing the number of swept MW frequencies by a factor of four. 

\begin{figure*}[htbp]
\begin{center}
\begin{overpic}[width=0.95\textwidth]{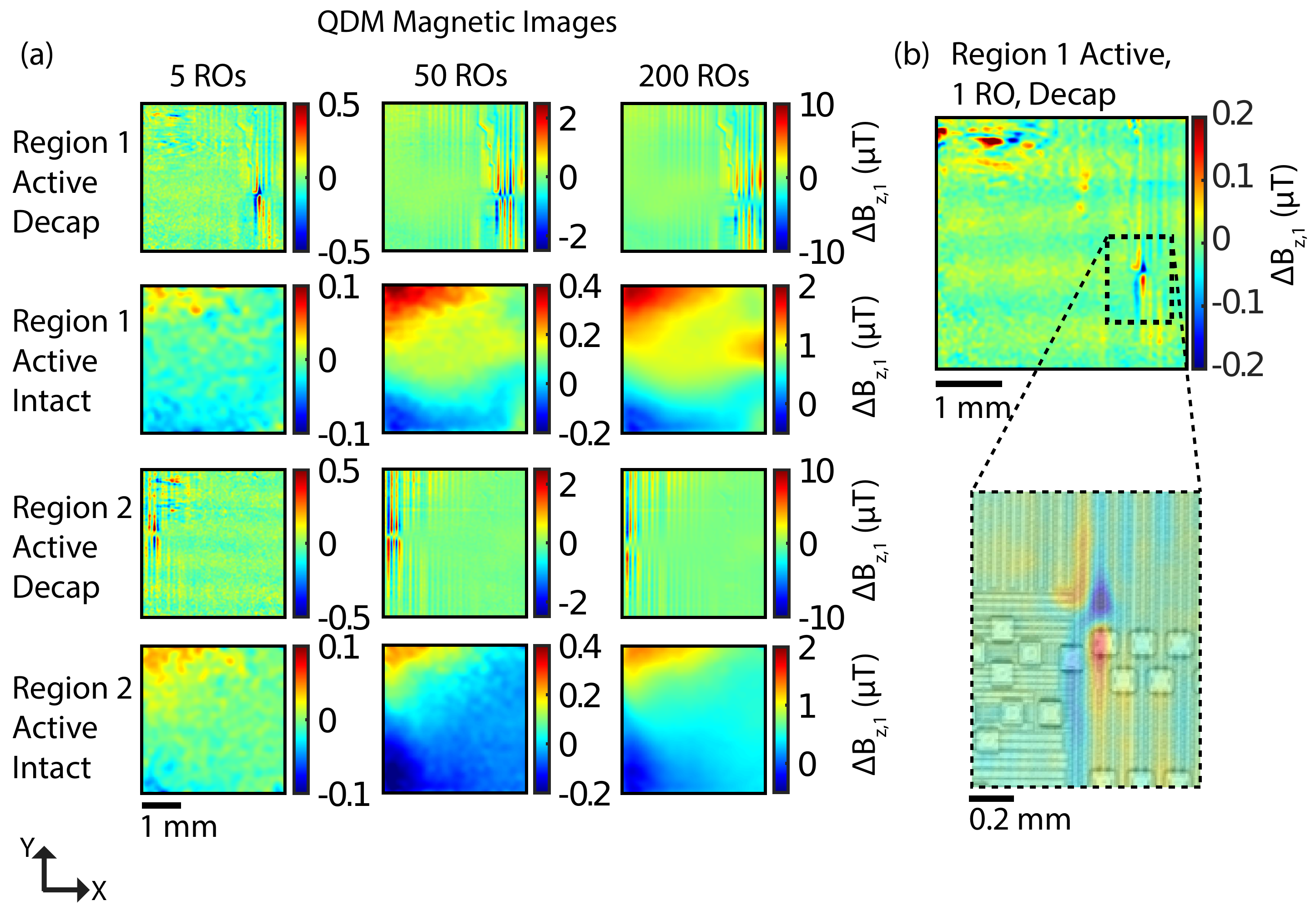}
\end{overpic}
\end{center}
\caption{\label{RONumber}
(a) QDM magnetic images indicate sensitivity to changing the number of ROs in different regions for the  decapsulated and intact chip when performing overlapped measurements. Different scale bars are used for feature clarity. (b) Decapsulated QDM data of $\Delta B_{z,1}$ for a single active RO in Region 1, demonstrating measurement sensitivity to the magnetic field from current supplying 1 RO. Inset: Overlay of the magnified single RO magnetic field image with a high resolution optical image of the circuit die. Each image is the average of 10 QDM (nominally identical) measurements.}
\end{figure*}

Fig.~\ref{RONumber}(a) shows example projection magnetic field images, averaged over ten measurements, for 5, 50 and 200 ROs in regions 1 and 2 for the decapsulated FPGA and the intact FPGA. Magnetic fields are generally reduced with diminishing numbers of ROs, due to the smaller current densities required for power distribution to smaller clusters. The maximum field amplitude is found not to scale linearly with number of ROs due to the currents being distributed over a differing number of wires on the top metal layer. The $\sim$ 200 nT magnetic field arising from a single RO is detectable for the decapsulated chip, Fig.~\ref{RONumber}(b), given the experimental noise floor of 20~nT (see Supplemental Material \cite{suppl}). The overlay of the measured magnetic field and the top metal layer illustrates potential location of vias where current is routed to deeper metal layers. 

Magnetic fields measured for the intact chip are decreased in magnitude and have lower intrinsic spatial resolution due to the large stand-off distance, compared to the decapsulated chip. The suppression of higher spatial frequency signals at large stand-off distances allows for more aggressive binning and spatial filtering of the intact data, without sacrificing spatial resolution and field information (see Supplemental Material \cite{suppl}). This approach enables a lower noise floor of 2~nT for the intact chip data, which partially overcomes the reduction of field amplitude with distance. For some regions of the field-of-view the noise floor is limited by state-independent variation in the magnetic field (See Supplemental Material \cite{suppl}) likely due to long-time power instability of the board. In order to enhance sensitivity and push the speed at which measurements can be taken, diamonds with thicker NV layers can be utilized at the cost of spatial resolution (see Supplemental Information \cite{suppl}). Such methodology may be especially beneficial when performing intact measurements, where the spatial resolution is already limited by the package stand-off distance. 

The dependence of current on the RO cluster size leads to state dependent temperature changes of the FPGA, which are determined from NV ODMR measurements using Eqn.~\ref{SolvedFields} Due to the high thermal conductivity of the diamond chip, there is no spatial structure in the resultant temperature maps. However, from temperature measurements over the entire field-of-view, we are able to determine a scaling of $\sim$ $.0075^{\circ} C$ per active ring oscillator (see Supplemental Material \cite{suppl}); with a temperature increase of $\sim$ $1.5^{\circ} C$ for the 200 RO state.

\subsection{Magnetic Field Source Interpretation}
The QDM magnetic field images shown in Figs. \ref{VectorField} and \ref{RONumber} result from current density sources located at various depths in the decapped and intact FPGA. Current is distributed in the interconnect layers of the silicon die and the package substrate. Each layer acts as a quasi two-dimensional current source contributing to the overall magnetic field detected by the NVs. The stand-off distance between the NV sensing plane and the current sources determines which metal layer dominates the field measurement. Generally, small wire features close to the sensing plane will dominate for small stand-off distance and large wire features far from the sensing plane will dominate for large stand-off distances.

For example, the 21~$\mu$m wide wires of the top metal layer contribute to the measured $\Delta B_X$ and $\Delta B_Z$ fields in the QDM images of the decapped FPGA, as seen by the spatial features of the fields in Fig.~\ref{VectorField}. Topside decapsulation removes the 500~$\mu$m of epoxy packaging above the die shown in the SEM in Fig.~\ref{Artix}(e). This results in a $5-10$ $\mu$m stand-off distance between the top metal layer and the NV sensing plane which is sufficiently small to resolve the spatial variation of fields resulting from currents in the top metal layer. Fields from smaller wires in the metal stacks below the top metal layer are too distant to contribute significantly to the measured field. 

The measured magnetic field distributions for both the decapsulated and intact chips include contributions from large current sources far from the NV sensing plane. These sources consist primarily of the metal layers of the 400~$\mu$m thick package substrate. The 300~$\mu$m silicon die separates the NV layer from the top of the package substrate for the decapsulated chip. An additional separation of $\sim$500~$\mu$m due to the epoxy gives a total stand-off distance of about 800~$\mu$m for the intact chip. These large current sources result in the broad features of the measured $\Delta B_Y$ data for the decapsulated chip in Fig.~\ref{VectorField}(a), and of the measured $\Delta B_{z,1}$ for the intact chip in Fig.~\ref{RONumber}(a). The dominant contribution of the substrate layers explains differences in the measured fields of the intact chip compared to those of the decapsulated chip, even when the latter are low pass filtered to account for the difference in measurement stand-off.

Comparison of the measured data with finite element analysis (FEA) simulations support the interpretation of the data as resulting from contributions of current sources in different layers at different depths from the NV plane. The FEA model, constructed in COMSOL Multiphysics, consists of 21.6~$\mu$m wires in the top metal layer with inter-wire spacing of 12.7~$\mu$m, and 100~$\mu$m thick metal wires in the package substrate layer with inter-wire spacing of 100~$\mu$m. An inter-layer separation of 300~$\mu$m represents the thickness of the silicon die. A current of $\sim$10~mA is applied to the wires in each layer with alternating bias to approximate the current of 200 active ROs. Plots of $B_Z$ for planes at 25~$\mu$m and a 500~$\mu$m above the top metal layer are given (see Supplemental Material \cite{suppl}) for comparison with the NV measurements at the nominal stand-off distances for decapsulated and intact chips respectively. The spatial features of the small wires are only evident in the $B_Z$ field of the plane with small stand-off, whereas the contribution of the large wires dominates at large stand-off distances.

The measurements presented in Figs. \ref{VectorField} and \ref{RONumber} are the net static magnetic fields resulting from steady-state RO operation in the FPGA. The static fields are interpreted to result from a time-averaged superposition of dynamic current draws from the top metal layer to the transistor level. The ROs used for this experiment each consist of three CMOS inverters that sequentially switch state during RO operation. A small, short-circuit current spike occurs in every inverter that switches state (due to simultaneous conduction through the two transistors of the inverter inducing a transient current path from supply voltage to ground). However, the individual switching of the inverters in the ROs is not temporally synchronized, resulting in a time-averaged, steady-state current draw from the top metal layer, and a consequently measurable static magnetic field (see Supplemental Material \cite{suppl}). 

\section{Analysis}

\begin{figure*}[ht]
\begin{center}
\begin{overpic}[width=0.95\textwidth]{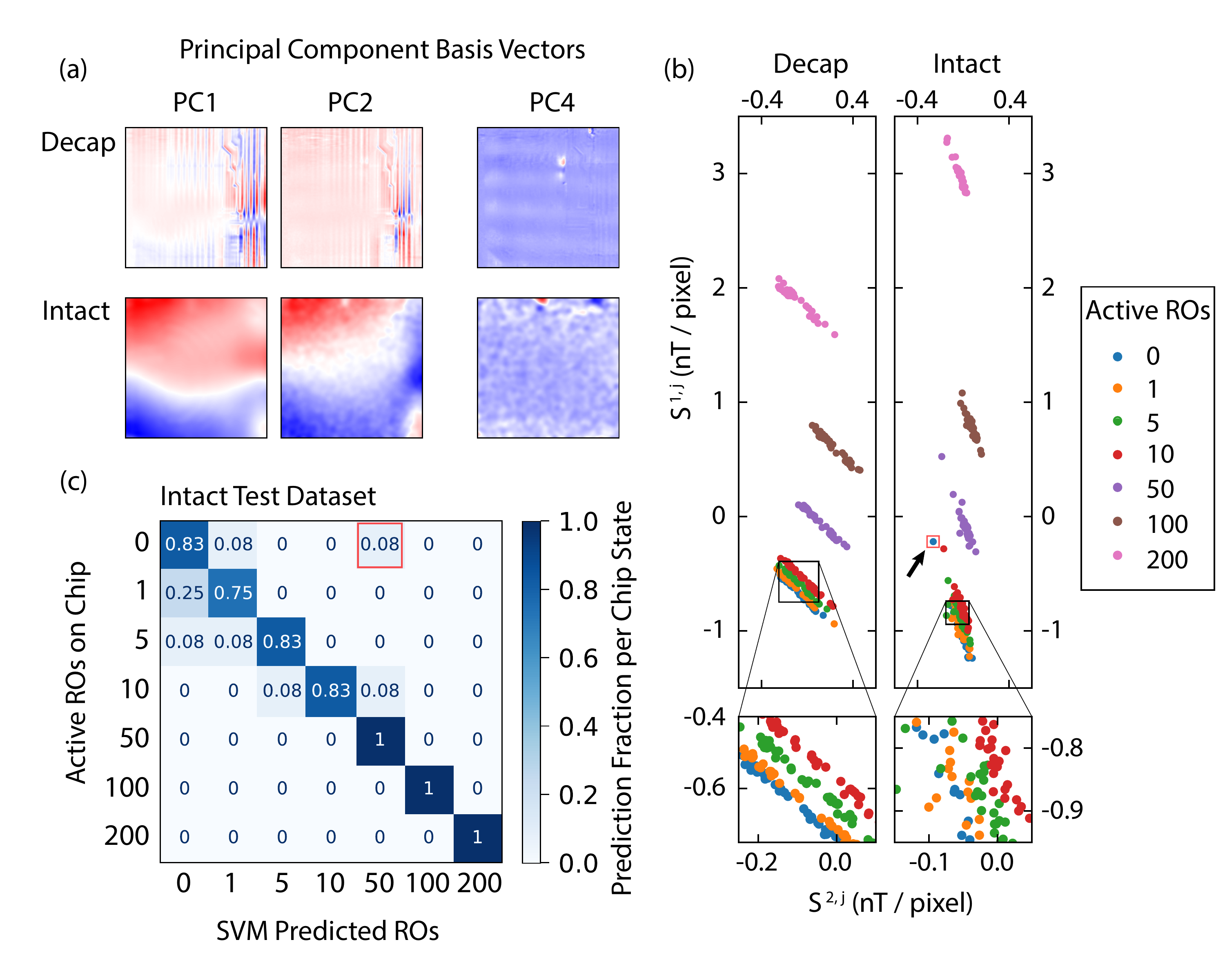}
\end{overpic}
\end{center}
    \caption{Principal component analysis (PCA) and support vector machine (SVM) classification of QDM images. Region 1 is active with 0, 1, 5, 10, 50, 100, or 200 ring oscillators (ROs). (a) Example principal component (PC) basis vectors plotted as images for both decapsulated (decap) and intact data sets. PC1 and PC2 are shown to exemplify PCs that resemble magnetic field images and thus will be useful in chip state classification. PC4 is shown as an example PC that captures activity state independent variations and thus will not be useful in chip state classification. (b) The PCA score for PC1, $S^{1,j}$, is plotted against the score for PC2, $S^{2,j}$, for each magnetic field image, $\mathbf{B}^j$, as a demonstration of state distinguishability . This distinguishability is evidenced by the separation of colors representing differing numbers of active ROs. Insets magnify the scores for small numbers of ROs, and show greater fidelity of state separation in the decapsulated data set compared to the intact data set. (c) Table of SVM predictions on the test set for the intact images. For each unique value of active ROs, there are 12 images in the test set. Rows indicate the fraction of images predicted for each of the possible chip states. All but one prediction (indicated by the red boxes in (b) and (c)) lie on or near the main diagonal demonstrating the high predictive power of the SVM classifier. The corresponding table for the decapsulated data set is not shown, as the main diagonal would contain 1's and the off diagonals would contain 0's due to the perfect separability of each state (see Table \ref{tbl:prediction_accuracy}).}
    \label{fig:pca_analysis}
\end{figure*}

Forward modeling of the current distributions and resulting magnetic fields for the different RO states programmed on the FPGA is an intractable problem without complete knowledge of the wire layout and current paths. Interpretation of the QDM measurements by comparison with forward-model simulations is therefore limited to the arguments such as those presented in the previous section. However, automated machine learning (ML) algorithms can be applied to the QDM data to discriminate between and ultimately classify the different operating states. Ideally, a magnetic field image is used as input to an ML algorithm, and the number of ROs is determined as the output. In practice, this problem is approached with a limited data set of magnetic field images for each FPGA state; and a dimensionality reduction algorithm is employed before applying a classification technique using the Python\cite{Python} package \texttt{scikit-learn}\cite{scikit-learn}.

\subsection{Data Preprocessing}
QDM data undergoes a series of preprocessing steps in preparation for dimensionality reduction and classification. Only images with Region 1 active are used so that the number of ROs is predicted by the classification scheme. The number of ROs activated for any given image is one of 0, 1, 5, 10, 50, 100, or 200. The data set consists of 40 QDM images per RO state for the decapsulated chip and 32 images per RO state for the intact chip. These $M \times N$ images are subsequently binned such that the decapsulated images contain $600 \times 606$ pixels and the intact images contain $300 \times 303$ pixels, while covering the same field-of-view. Measurements of the idle state (0 ROs) are randomly taken during data collection to account for long term drifts. These idle state measurements are subtracted from active state images nearest in time. The intact and decapsulated data sets are split into training and test sets so that the prediction accuracy of the trained model can be estimated on data that the training procedure has not encountered. The splits are 75\%/25\% for the decapsulated images and 64\%/36\% for the intact images. 

\subsection{QDM Image Dimensionality Reduction}
Each magnetic field image is composed of $\sim 10^5$ pixels and thus occupy a high dimensional space for classification. Principal component analysis \cite{PCA_1,PCA_2} (PCA) is therefore used to reduce the dimensionality of the classification problem. PCA is a well-established technique that determines the highest variability axes of a high-dimensional data set. PCA amounts to an eigenanalysis where the eigenvectors, called principal components (PC), correspond to the axes of interest; and the eigenvalues relate to the amount of data variance along the respective PCs.

PCA is applied separately to the data sets of the decapsulated chip and the intact chip with the \texttt{scikit-learn} class \texttt{decomposition.PCA()} and yields PCs such as those plotted in Fig.~\ref{fig:pca_analysis}(a). Spatial patterns evident in the PCs are also present in the magnetic field images of Fig.~\ref{RONumber}(a), confirming that these features are physically significant and important for distinguishing between different samples. There exist as many PCs as dimensions in the data set; however, only the first several PCs capture non-noise based information (see Supplemental Material \cite{suppl}). We determine that $>$99$\%$ of the variance in the intact and decapsulated data sets is captured by the first 9 PCs, which are therefore the only PCs used in this analysis. 

The scores of these first 9 PCs are used to effectively reduce the dimensionality of the magnetic field images from $\sim 10^5$ pixels to 9 scores. The PC scores, $S^{i,j}$, are determined by taking the dot product of the $i^{\text{th}}$ PC, defined as $\mathbf{W}^i$, with the $j^{\text{th}}$ image, $\mathbf{B}^j$, and normalized by the total number of pixels. This gives 
\begin{equation}
    S^{i,j}=\frac{1}{MN}\sum_{m=1}^M\sum_{n=1}^N W_{m,n}^i B_{m,n}^j
    \label{PCScore}
\end{equation}
for the first 9 PCs. Fig.~\ref{fig:pca_analysis}(b) shows examples of the PCA scores: the score for PC1 is plotted against the score for PC2 for each magnetic field image of both the decapsulated and intact data (additional principal components and score plots are given in Supplemental Material \cite{suppl}). The plot is color coded by number of active ROs showing that these two scores are useful in distinguishing the number of active ROs on the FPGA for both decapsulated and intact measurements. Classification of the active number of ROs is accomplished by using the first 9 PCA scores as input to a support vector machine (SVM) classifier algorithm. The spread of data points along a fixed slope for each state in Fig.~\ref{fig:pca_analysis}(b) is consistent with small offsets between different image acquisitions (see Supplemental Material \cite{suppl}).

\subsection{IC Activity State Classification}
A support vector machine \cite{SVM_1} (SVM) is the supervised classification technique used to classify the magnetic field images, leveraging their key features characterized by the PCA scores. SVMs are a set of algorithms that seek to classify samples by creating a boundary between categories of a training data set that maximizes the gap separating those categories. Samples from a test set are then classified in relation to this boundary. 

The \texttt{scikit-learn} class \texttt{svm.SVC()} is used as a multi-dimension, multi-category classifier. The categories for classification are the chip states given by the number of ROs. The dimensionality is given by the 9 PCA scores recorded for each image. PCA scores are fit to the known FPGA states with a linear SVM model and a regularization parameter of $C=6$ (see Supplemental material \cite{suppl}). 

\subsection{Classification Results}

\begin{table}
\centering
\begin{tabular}{c | c c c c c c c | c}
    \hline
    & \multicolumn{7}{c|}{Number of ROs (Region 1)} & \multirow{2}{*}{Total} \\
    & 0 & 1 & 5 & 10 & 50 & 100 & 200 \\ \hline
    Decapsulated & 1.00 & 1.00 & 1.00 & 1.00 & 1.00 & 1.00 & 1.00 & 1.00 \\
    Intact & 0.83 & 0.75 & 0.83 & 0.83 & 1.00 & 1.00 & 1.00 & 0.89 \\
    \hline
\end{tabular}
\caption{Chip state prediction accuracy on the test dataset.}
\label{tbl:prediction_accuracy}
\end{table}

The full ML model, including preprocessing, PCA, and SVM, is fit using the training set and subsequently evaluated on the test set, for both decapped and intact FPGA data. A prior step of cross-validating the model hyperparameters is taken for the intact FPGA data set (Supplemental material \cite{suppl}). The ML model efficacy, summarized in Table \ref{tbl:prediction_accuracy}, is determined by the accuracy of the test set evaluations. FPGA activity states are well separated in PCA-space for the decapsulated data set. Predictions on the test set consequently yield perfect accuracy, even for small numbers of ROs, consistent with expectations (Supplemental material \cite{suppl}).

Results for intact data set are similarly well separated for large numbers of ROs, resulting in perfect prediction accuracy for $\geq$50 ROs. However, FPGA activity states are not fully separated for $<$50 ROs, resulting in imperfect predictions. Nonetheless, the trained ML model  achieves $\sim$ 80\% accuracy for each of 0, 1, 5, and 10 RO active states. Fig.~\ref{fig:pca_analysis}(c) additionally shows that incorrect predictions are nearly always close to the correct state. 
For example, the model predicts 5 ROs correctly in 83$\%$ of test cases, with misclassifications of 0 or 1 RO otherwise. The red box in Fig.~\ref{fig:pca_analysis}(c) indicates a single case for which the classifier incorrectly predicts 50 ROs for a 0 RO state. An arrow and analogous red box in Fig.~\ref{fig:pca_analysis}(b) shows that the PCA score for this state is an outlier in the data. The positive classification results presented in Fig.~\ref{fig:pca_analysis} give an initial demonstration of the capability of ML techniques to identify chip activity via non-invasive measurements with the QDM. Larger data sets populating ML methods have promise to enable classification of a wide array of chip activity in the context of hardware security and fault detection. 

\section{Outlook}
We present the first demonstration of NV diamond imaging of the static (DC) magnetic field emanations from an FPGA. The ensemble NV measurement technique of the Quantum Diamond Microscope (QDM) yields simultaneous wide field-of-view (few mm) with high resolution ($\sim 10 \mu$m) vector magnetic field images, which is  not achievable using other techniques. We further demonstrate that these images can be used with machine learning (ML) techniques to quantifiably determine the active state of the FPGA integrated circuit (IC), for both decapsulated and intact chips. The  fidelity of classification is dependent on the activity level and stand-off distance between the circuit currents and the measurement plane. Our results show conclusively that it is possible to use static magnetic field measurements to identify targeted, active states on an FPGA without requiring time domain data. 

QDM magnetic imaging of ICs has significant potential to augment state-of-the-art diagnostic techniques in areas ranging from fault detection, Trojan detection, counterfeit detection, watermarking, and electromagnetic side channel characterization. We show that patterns of DC magnetic fields from the power distribution network of an IC are a powerful indicator of chip activity. The simultaneous wide-field and high-resolution magnetic imaging provide by the QDM may further enable detection of correlated spatial and temporal (e.g., transient) events over sequential measurements that is not possible with scanning magnetometry techniques.

Technical improvements enabling faster QDM measurements \cite{qdmEdlyn, barry2019sensitivity} will be implemented in future work for extended data collection and further analysis techniques. Larger data sets will also allow for leveraging the full power of convolutional neural networks for advanced state classification. Further time-resolved measurements will also allow for separation of magnetic fields by temporal or frequency profiles. It is possible such measurements could be used to resolve the magnetic fields specific to adjacent circuitry, for example from clock and power distribution networks as well as gate level activity, proving further indication of chip activity. The unique capabilities of NV diamond magnetic field imaging show great promise for integrated circuit applications and will be increasingly leveraged with ongoing development. 

\begin{acknowledgments}
We thank Ben Le for FPGA development; Adam Woodbury, Jeff Hamalainen, Maitreyi Ashok, Connor Hart, David Phillips, Greg Lin, CNS staff, Rebecca Cheng, and Amirhassan Shams-Ansari for helpful discussions and support; Raisa Trubko and Roger Fu for assistance on early measurements; Patrick Scheidegger and Raisa Trubko for work applying GPUfit to the analysis; Edward Soucy, Brett Graham, and the Harvard Center for Brain Science for technical support and fabrication assistance. This project was fully funded by the MITRE Corporation through the MITRE Innovation Program. P.K. acknowledges support from the Sandia National Laboratories Truman Fellowship Program, which is funded by the Laboratory Directed Research and Development (LDRD) Program at Sandia National Laboratories. NV-diamond sensitivity optimization pertinent to this work was developed under the DARPA DRINQS program (award D18AC00033). This work was performed in part at the Center for Nanoscale Systems (CNS), a member of the National Nanotechnology Coordinated Infrastructure Network (NNCI), which is supported by the National Science Foundation under NSF award no. 1541959. CNS is part of Harvard University. 
\end{acknowledgments}

\bibliography{ICrefs}

\end{document}